# Visualization and Travel Time Extraction System for the Statistics of TDCS Travel using MapReduce Framework

Eko Prasetyo
*Department of Informatics Engineering, Universitas Muhammadiyah Yogyakarta, Indonesia*
*eko.prasetyo@umy.ac.id*

Prayitno
*Department of Electrical Engineering, Politeknik Negeri Semarang, Indonesia, prayitno@polines.ac.id*

Jing-Doo Wang
*Department of Computer Science and Information Engineering, Asia University, Taiwan,*
*jdwang@asia.edu.tw*

Karisma Trinanda Putra
*Department of Electrical Engineering, Universitas Muhammadiyah Yogyakarta, Indonesia,*
*karisma@ft.umy.ac.id*

## Abstract

*Recently, extracting some information as a knowledge from big data is very challenging activity. The size of data is very huge and it requires some special techniques and adequate processing hardware. It is also applied in vehicles transportation data at Taiwan National Freeway from the Traffic Data Collection System (TDCS). The results of this extraction will be very useful if it can be used by the community. So that the delivery of information extracted from large data that is easily understood becomes a necessary thing. Presentation of results using images / visuals will make it easier for people to interpret the information provided. In this project, an interactive visualization of the results of extracting statistical information is attempted to be provided. The results can be used by users to support the decision making of road users in determining the appropriate time when going through the road pieces around the Taichung City. This visualization of the statistics will help people who want to predict the travel time around Taichung City.*

**Keywords**: big data; Map Reduce; interactive visualization; travel time.

**JEL Classification:** C80: Data Collection and Data Estimation Methodology • Computer Programs/General

# 1. Introduction

Extracting daily traffic features and information in certain area become interesting in big data era. Traffic extraction pattern become important at many levels of transportation planning and management. This information, information about link travel pattern can reveal problematic locations where targeted mechanism can be introduced to improve service ability performance. On the other hand, this information can be useful for commuters or daily traveler as urban mobility to better understand in traffic roads, since it can aid them about travel decision (Ma, Koutsopoulos, Ferreira, & Mesbah, 2017). Furthermore, traffic extraction pattern can be useful step to reduce traffic congestion. One of the examples suffered from traffic congestion was in Boston, USA. It was reported that the traffic congestion cost US $87 billion in lost productivity in 2018 and will increase each year[1].

In big data era information, we drowning in a lot of data but starving for knowledge. One of the best examples is real-time data generated from high-speed highway Road Network Traffic Management System and Highway Electronic Toll Collection System at Taiwan National Freeway[2]. This vehicle transportation database system recorded every car, bus and truck passes at highway road. Each vehicle data that pass in gate, pass out gate, kilometers run and toll fee collected in this system. It is important to note that, each day the system will give a 200 MB data, so it will be more than 25 GB data every month. It is believed that, such big data processing could give significant knowledge into transportation system. Furthermore, the ability to predict traffic information based on big open data is one of important steps to gain the effectiveness of traffic control system.

Using large historic data and real-time traffic, researchers proposed many methods to provide patterns, knowledge, predict and forecasting vehicles traffic pattern. Wang et. al. proposed a method to extract significant patterns of time interval from Taiwan Freeway Gantry Timestamps sequence using Hadoop MapReduce framework (Li, Ooi, Özsu, & Wu, 2014; J.-D. Wang & Hwang, 2017). The MapReduce Framework was proposed by Google become popular as parallel and distributed computing to process very large data (Dean & Ghemawat, 2008). Traffic data contains several types like timestamp, vehicles type, and fees. Previous researchers use MapReduce framework to solve imbalance big data using Random Forest method (Del Río, López, Benítez, & Herrera, 2014) that traditional data mining approaches are not able to cope with new requirements imposed by bigdata.

In this paper we propose vehicle traffic pattern extraction using two methods. First, our approach is designed parallel computation and distributed Taiwan open freeway data by utilizing MapReduce program. Second, we visualize the traffic pattern and analyze the time travel.

---

[1] https://www.cnbc.com/2019/02/11/americas-87-billion-traffic-jam-ranks-boston-and-dc-as-worst-in-us.html
[2] http://tisvcloud.freeway.gov.tw/

The main contributions of the paper are:

- We construct a method to extract significant vehicle time travel interval based on Taiwan national Freeway open data.
- We conduct an exploratory analysis of vehicle time travel interval.

The rest of this paper is organized as follows. In section 2, we introduce a literature review to support our problem solution. Furthermore, we proposed our methods extraction using MapReduce framework in section 3. Then in section 4, we carry experiments and launch analysis of the result. Finally, we conclude the paper result and discuss a future work in section 5.

## 2. Literature Review
### 2.1. Time interval extraction

Recently an Intelligent Transportation System (ITS) has gain attention to the researchers. This system conducts several areas such as integrated application of communications, control, and information processing on the transportation sector. Data source feed to ITS come from several devices namely road sensors, camera traffic light or vehicles sensors (Imawan, Putri, An, Jeong, & Kwon, 2015; Prayitno, Ardjo, Triyono, & Kuswanto, 2019). Extracting information from these data feed could be benefit for the government and commuter. (J.-D. Wang & Hwang, 2017) illustrate the travel time extraction in Taiwanese government freeway open data platform using MapReduce framework. The information extracted from the data give a useful insight about the longest trip within Freeway No.5 both in southern and northern direction. In addition, (Chang, Chen, Hsu, & Yang, 2016) extracted Electronic Toll Collection (ETC) data based on seamless-temporal data fusion to improve travel time prediction accuracy of different locations and times. Developed using Java programming the experiment shows 10% considered as highly accurate predictions.

Machine learning and data mining methods used by several researchers to extract time interval and predict the vehicles traffic. (Zhang, Liu, Yang, Wei, & Dong, 2013) used an improved K-nearest neighbor model to predict the vehicles traffic flow. The method is to predict short-term urban expressway flow and the result show over 90 percent describe the feasibility of the methods. Furthermore, (Xiaoyu, Yisheng, & Siyu, 2013) also conducted traffic flow forecasting using K-nearest neighbor algorithm because it does not require a priori knowledge and perform better forecasting in linear model. Another researcher (Charlotte, Helene, & Sandra, 2017) extracted database of accidents, major road works and traffic loop detectors on 23 urban highways. It extracted explanatory variables such as the travel time distribution, number of lanes, time of the day, direction number of accidents, state of repair of the road and road works. In addition, estimating travel times and vehicles trajectories on freeways using dual loop detectors conducted by (Coifman, 2002) described link travel time is considered to be more informative to users than flow, velocity, or occupancy measured at a point detector. The accuracy of the method lends further evidence that the linear approximation of the flow density relationship is reasonable during congestion. As traffic flow is complicated in nature, deep learning algorithm can represent traffic features without prior knowledge, which has good performance for traffic flow prediction (Lv, Duan, Kang, Li, & Wang, 2014). The deep learning method approach with SAE model for traffic flow prediction successfully discover the latent traffic flow feature representation such as non-linear spatial and temporal correlations from the traffic data. Another variable should be considered to examine the effect of information on traveler's perception, accuracy/speed of comprehension, and mental workload. (Saedi & Khademi, 2019) noted that information on traveler's cognition, recall ability, and persistence of data in memory could enhance travel time extraction in intelligence transport system.

## 2.2. MapReduce

MapReduce become important framework in bigdata processing. The framework is a processing technique and program model which process large amount of data (multi-terabyte data-sets) in-parallel on large cluster (thousands of nodes) of commodity hardware in a reliable, fault-tolerance manner. The MapReduce algorithm contains two important tasks namely Map and Reduce as shown in Figure 1. Map takes a set of data and convert it into another set of data, where individual elements are broken down into tuples (key/value pair). A key-value pair is a basic data structure for an input and output of the MapReduce framework. Furthermore, reduce task takes the output from a map as input and combines those tuples into a smaller set of tuples. As the sequence of the name MapReduce implies, the reduce task is always performed after the map task.

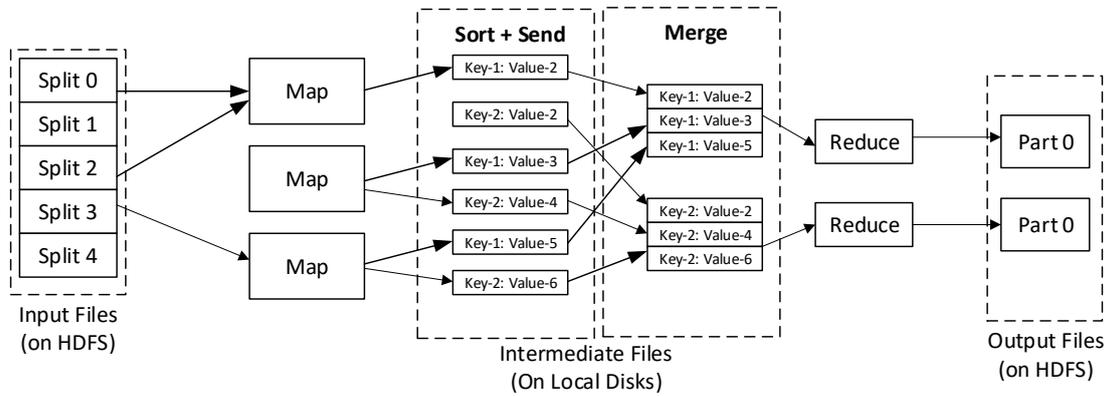

Figure 1. MapReduce Framework

(Xia, Wang, Li, Li, & Zhang, 2016) described a method using MapReduce framework on a Hadoop platform to developed a distributed spatial-temporal weighted model. The method is to forecasting a traffic flow in Bigdata ecosystems using KNN computing. Furthermore, for big data processing MapReduce framework effectively extracted significant pattern histories from timestamped text (J. D. Wang, 2016).

# 3. Methodology
## 3.1. Methods

Figure 2. shown the flowchart to process data extraction and show statistic of vehicles distribution. First process is to upload TDCS data source form (J.-D. Wang & Hwang, 2017) to Hadoop File System. Then, we define month data interval (1-month, 3 month or 6 month). Next step was to choose start gantry and end gantry. Furthermore, at the next process Hadoop MapReduce framework will extract vehicles traffic distribution at specific Freeway intersection. Finally, we present and describe statistics about vehicles traffic distribution.

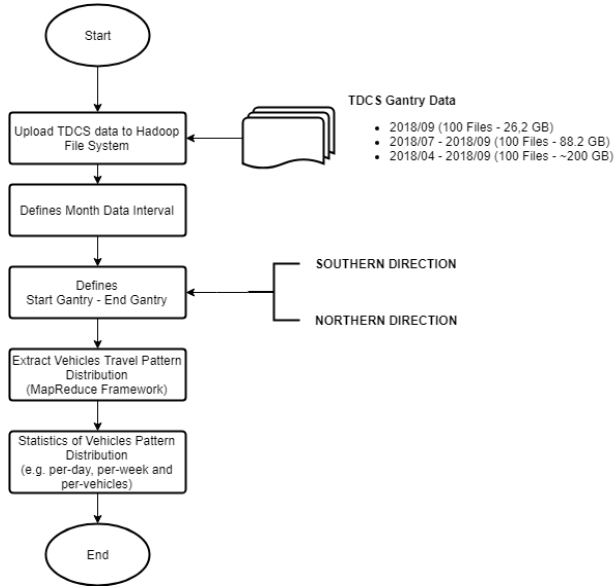

**Figure 2.** The flowchart of Information Extraction at Freeway Intersection and computing statistics of vehicle distribution

## 3.2. Gantry Timestamp sequence from Traffic Data Collection System (TDCS)

Data vehicle traffic was retrieved from Taiwan Data Collection System (TDCS). TDCS as one of open data platform system provide by Taiwan Government. Table 1 shows data collection for experimental purpose in this report.

**Table 1.** Input Data Set TDCS

| Data Set | Number of Month | Number of Files | The Total File Size (GB) |
|---|---|---|---|
| 2018/09 | 1 | 100 | 26.2 |
| 2018/07 – 2018/09 | 3 | 100 | 88.2 |
| 2018/04 – 2018/09 | 6 | 100 | ~ 200 |

This data source is from Wang et.al research (J.-D. Wang & Hwang, 2017) that was processed TDCS Daily data using MapReduce Framework. These data ranging from 1 month (2018/09) to 6 months (2018/04 – 2018/09 stored information about vehicles timestamps at each gantry at Freeway. The total size of each file almost 26 GB size and 100 files each month. Furthermore, total size of all 3 months data experiments is 88 GB and 100 files, to extract information from this huge data at one standalone computer will take a long time. MapReduce framework as parallel and distributed computing is one of solution to compute this data.

### 3.3. Extracting Daily Traffic Interval Pattern

To extract vehicles distribution at Taiwan National Freeway 01 and 03 at Taichung city, we must define the start gantry and end gantry for each travel. For both national freeways we extract pattern travel from southern and northern direction. For national freeway 01 we choose gantry 01F-157.2N and 01F-180.2N for northern direction, then gantry 01F-180.2S to 01F-157.2S for southern direction as shown at Table 2. Gantries Description South and North Direction (National Taiwan Freeway 01 at Taichung city). Furthermore, Figure 3 depict the gantries location for south and north direction.

Table 2. Gantries Description South and North Direction (National Taiwan Freeway 01 at Taichung city)

| Direction | | Gantry ID | Distance (KM) | Fee (TWD) | Interchange (start) | Interchange (stop) |
|---|---|---|---|---|---|---|
| Northern | 1 | 01F-157.2N | 10.5 | 18.9 | hòu lǐ 后里 | sānyì 三義 |
| | 2 | 01F-162.1N | 4.7 | 8.4 | Taichung system 台中系統(連接國4) | hòu lǐ 后里 |
| | 3 | 01F-166.4N | 2.5 | 4.5 | Fengyuan 豐原 | Taichung system (connecting country 4) 台中系統(連接國4) |
| | 4 | 01F-172.5N | 6.2 | 9.3 | Taiya 大雅 | Fengyuan 豐原 |
| | 5 | 01F-177.4N | 4.4 | 7.9 | Taichung (Taiwan Avenue) 台中(台灣大道) | Taiya 大雅 |
| | 6 | 01F-180.2N | 2.8 | 5 | Nanxun 南屯 | Taichung (Taiwan Avenue) 台中(台灣大道) |
| Southern | 1 | 01F-157.2S | 10.5 | 18.9 | sānyì 三義 | hòu lǐ 后里 |
| | 2 | 01F-162.1S | 4.7 | 8.4 | hòu lǐ 后里 | Taichung system 台中系統(連接國4) |
| | 3 | 01F-166.4S | 2.5 | 4.5 | Taichung system 台中系統(連接國4) | Fengyuan 豐原 |
| | 4 | 01F-172.5S | 6.2 | 9.3 | Fengyuan 豐原 | Taiya 大雅 |
| | 5 | 01F-177.4S | 4.4 | 7.9 | Taiya 大雅 | Taichung (Taiwan Avenue) 台中(台灣大道) |
| | 6 | 01F-180.2S | 2.8 | 5 | Taichung (Taiwan Avenue) 台中(台灣大道) | Nanxun 南屯 |

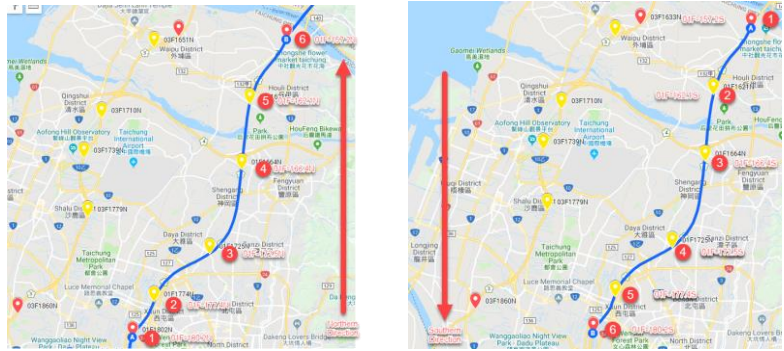

(a) Travel Pattern from 01F-180.2N to 01F-157.2N
(b) Travel Pattern from 01F-180.2S to 01F-157.2S
Figure 3. Travel Pattern from National Freeway 01 at Taichung city

Another travel pattern extraction from Taiwan National Freeway 03 at Taichung city start from gantry 03F-186.0N to 03F-163.3N for northern direction. On the other side, for southern direction are gantries id 03F-163.3S to 03F-186.0S as shown at Table 3.

Table 3. Gantries Description South and North Direction (National Taiwan Freeway 03 at Taichung city)

| Direction | | Gantry ID | Distance (KM) | Fee (TWD) | Interchange (start) | Interchange (stop) |
|---|---|---|---|---|---|---|
| **Northern** | 1 | 03F-186.0N | **8.8** | **15.8** | Héměi<br>和美 | Lóngjǐng<br>龍井 |
| | 2 | 03F-177.9N | **6.7** | **12** | Lóngjǐng<br>龍井 | Shā lù<br>沙鹿 |
| | 3 | 03F-173.9N | **3.7** | **6.6** | Shā lù<br>沙鹿 | Qīngshuǐ fúwù qū<br>清水服務區 |
| | 4 | 03F-171.0N | **3.5** | **6.3** | Qīngshuǐ fúwù qū<br>清水服務區 | Zhōnggǎng xìtǒng (liánjiē guó 4)<br>中港系統(連接國4) |
| | 5 | 03F-165.1N | **4.7** | **8.4** | Zhōnggǎng xìtǒng (liánjiē guó 4)<br>中港系統(連接國4) | Dà jiǎ<br>大甲 |
| | 6 | 03F-163.3N | **7.5** | **13.5** | Dà jiǎ<br>大甲 | Yuàn lǐ<br>苑裡 |
| **Southern** | 1 | 03F-163.3S | **7.5** | **13.5** | Yuàn lǐ<br>苑裡 | Dà jiǎ<br>大甲 |
| | 2 | 03F-165.1S | **4.7** | **8.4** | Dà jiǎ<br>大甲 | Zhōnggǎng xìtǒng (liánjiē guó 4)<br>中港系統(連接國4) |
| | 3 | 03F-171.0S | **3.5** | **6.3** | Zhōnggǎng xìtǒng (liánjiē guó 4)<br>中港系統(連接國4) | Qīngshuǐ fúwù qū<br>清水服務區 |
| | 4 | 03F-173.9S | **3.7** | **6.6** | Qīngshuǐ fúwù qū<br>清水服務區 | Shā lù<br>沙鹿 |
| | 5 | 03F-177.9S | **6.7** | **12** | Shā lù<br>沙鹿 | Lóngjǐng<br>龍井 |
| | 6 | 03F-186.0S | 8.8 | 15.8 | Lóngjǐng<br>龍井 | Héměi<br>和美 |

There are 6 gantries location from southern and northern direction shown at Figure 4. Travel Pattern from National Freeway 03 at Taichung city that extracted for vehicles travel distribution.

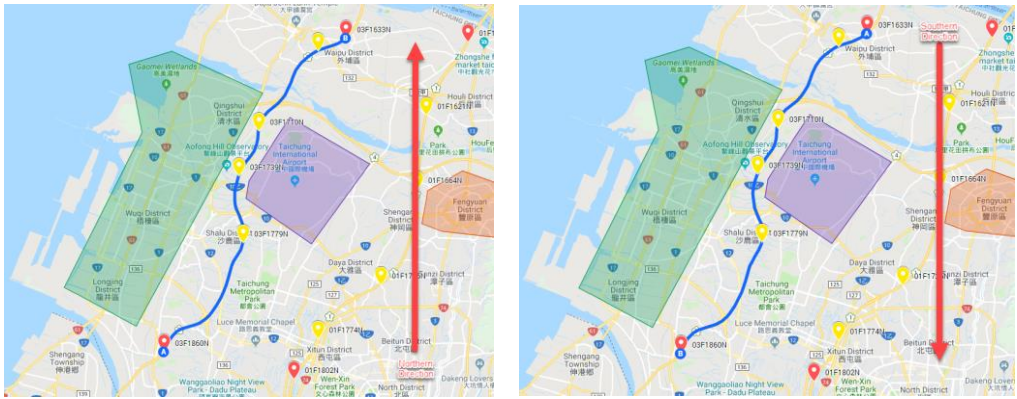

(a) Travel Pattern from 03F-186.0N to 03F-163.3N
(b) Travel Pattern from 03F-163.3S to 03F-186.0S
Figure 4. Travel Pattern from National Freeway 03 at Taichung city

# 4. Result & Analysis

The experiment result from gantries data extraction using Hadoop MapReduce framework. The result discussed distribution of traffic from 24 hours seven days at 1 months 2018/09. At section 3.1. extracted about number of traffic passed the Taiwan National Freeway 01 and 03 at Taichung city. Furthermore, at section 3.2. describe about vehicle type passed. It is important to note that in September 2018, there are 5 Saturday and 5 Sunday so vehicles distribution will be higher on these days' compare than the other days. Furthermore, 24 September 2018 is public holiday for Taiwan vehicles distribution also high around this day.

### 4.1. Traffic distribution National Freeway 01 vs National Freeway 03

Freeway traffic distribution can be seen from both southern and northern directions. Figure 5 depict the distribution vehicles from National Freeway 03 at Taichung city. To take a closer view, vehicles distribution peaked at 16.00 in the afternoon for both directions. It is important to note that, traffic distribution slowly increased from 06.00 to 10.00 in the morning. Furthermore, summation vehicles frequency to southern direction is greater than northern direction.

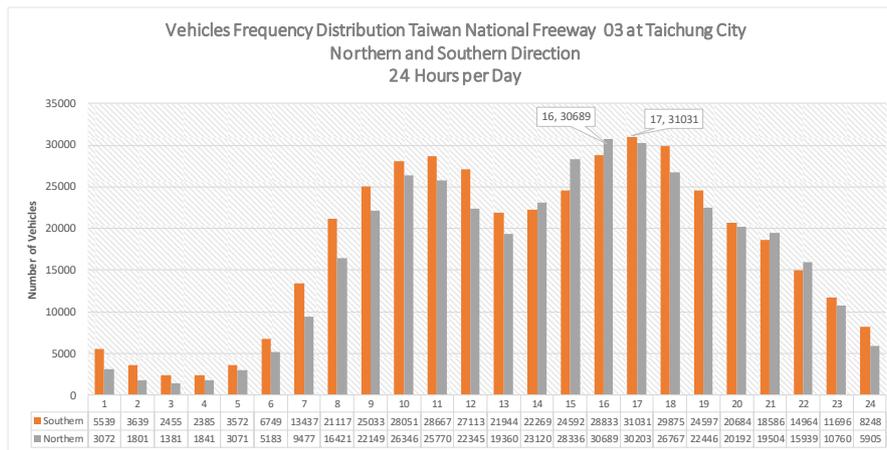

Figure 5. Traffic distribution at northern and southern direction at Taiwan National Freeway 03 at Taichung City from 24 hours

On the other hand, Figure 6. Illustrate Taiwan National Freeway 01 vehicles distribution. Summation vehicles frequency to northern direction is greater than southern direction. At this pattern vehicles distribution peaked at 10.00 a.m. in the morning for both directions. Vehicles traffic distribution for Taiwan National Freeway 01 and 03 are almost the same pattern.

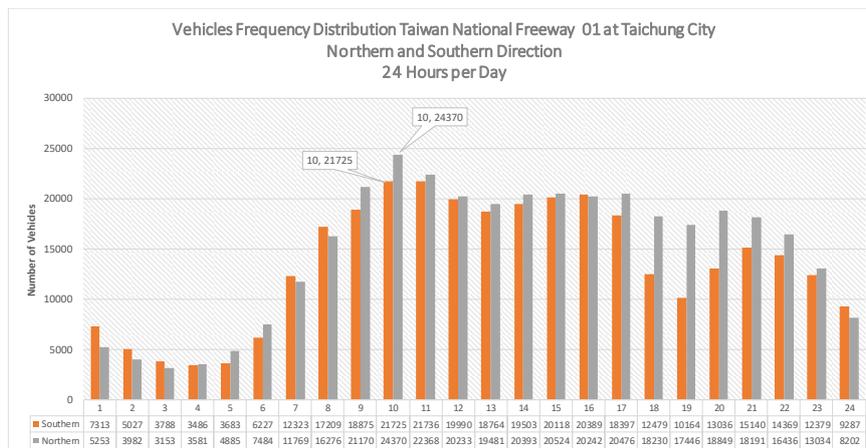

Figure 6. Traffic distribution at northern and southern direction at Taiwan National Freeway 01 at Taichung City from 24 hours

National freeway 03 is busier than national freeway 01, this can be seen in total number of vehicles pass through this freeway. National freeway 01 total number of vehicles pass by is 681,435 on the other hand national freeway 03 is 817,154.

### 4.2. Vehicle distribution frequency national freeway 01 vs national freeway 03 per week

Figure 7 and Figure 8 depicts the vehicles frequency distribution from Taiwan National Freeway 01 and National Freeway 03 respectively. It is interesting to take note that around 15.00 to 16.00 vehicles distribution at national freeway 03 become doubles than national freeway 01. Total number vehicles traffic distribution at "Sat" and "Sun" most affect the distribution. Most of people using their vehicles at 10.00 to 12.00 afternoon passed these freeways.

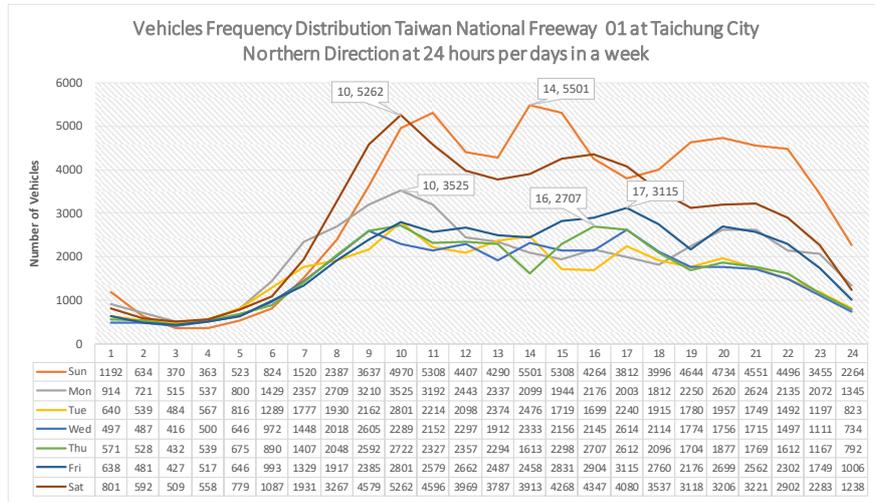

Figure 7. Traffic Distribution National Freeway 01 at Taichung city northern direction at 24 hours - seven days

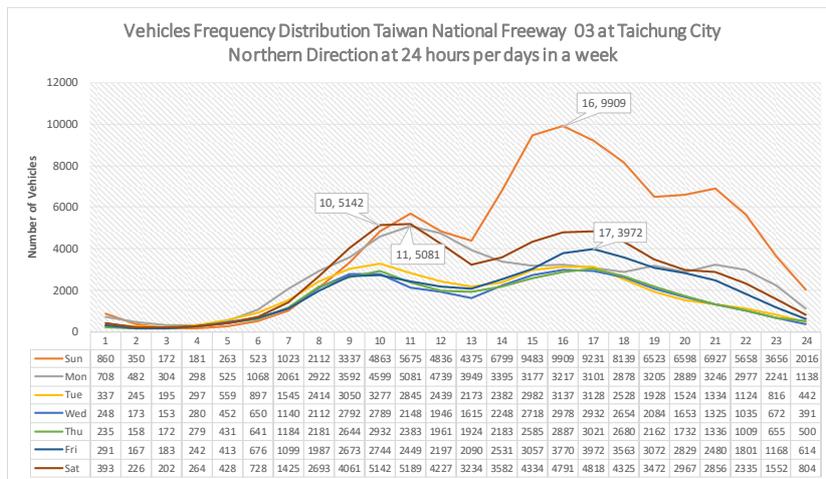

Figure 8. Traffic Distribution National Freeway 03 at Taichung city northern direction at 24 hours - seven days

On the other hand, national freeway 01 and national freeway 03 southern direction shown at Figure 9 and Figure 10 respectively.

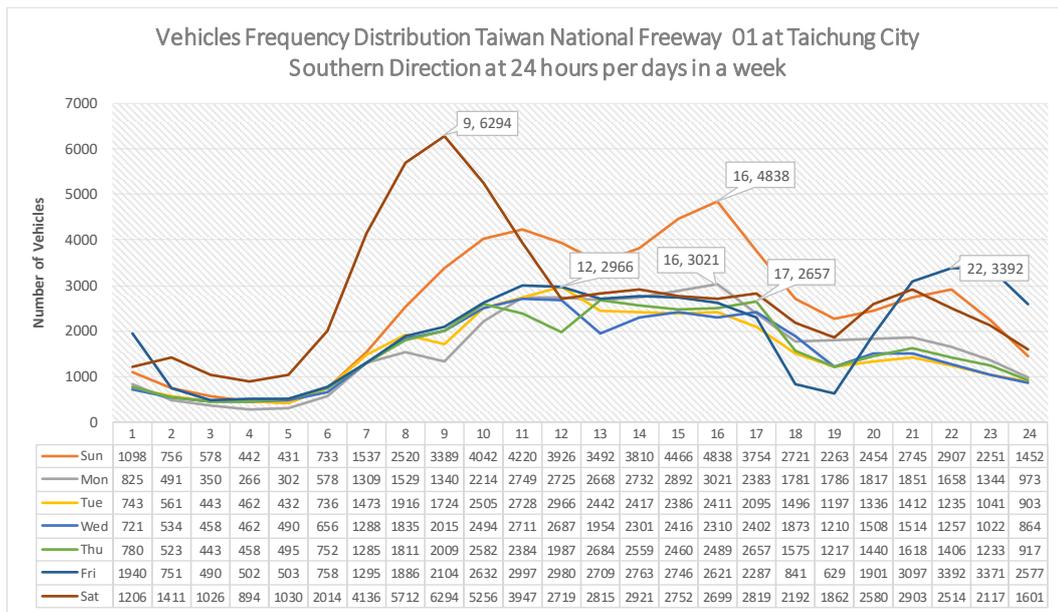

Figure 9. Traffic Distribution National Freeway 01 at Taichung city southern direction at 24 hours - seven days

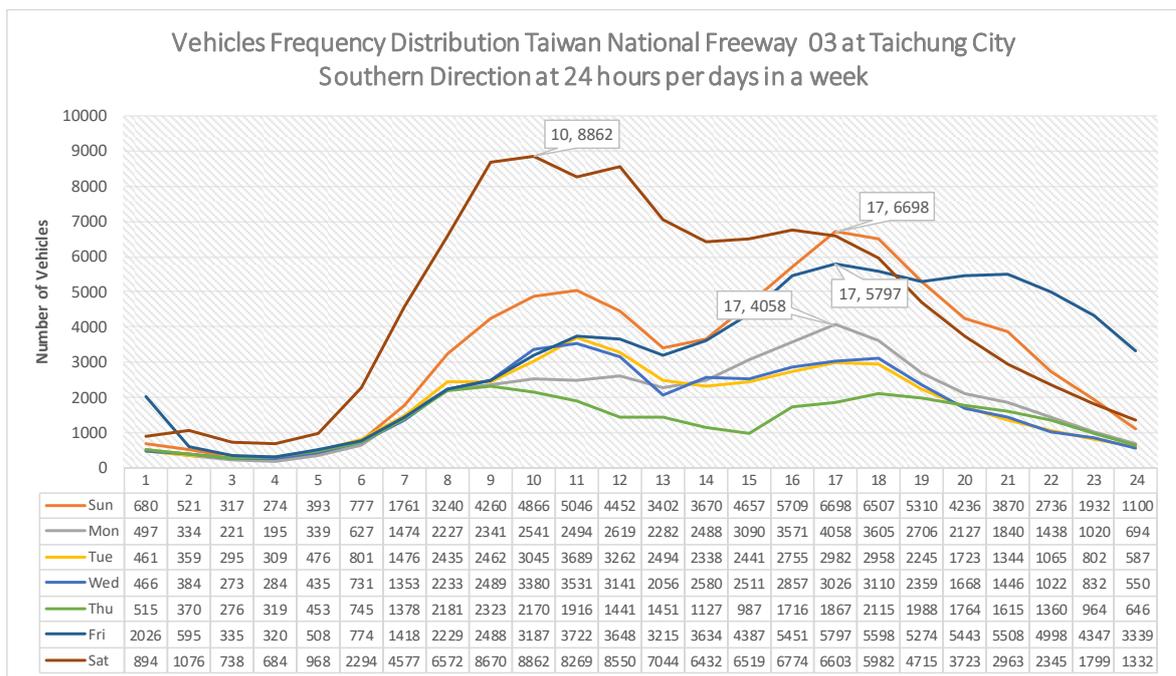

Figure 10. Traffic Distribution National Freeway 03 at Taichung city southern direction at 24 hours

## 4.3. Vehicle Type Traffic Distribution per week

There are several vehicle types recorded at gantry data, coded as "5" (Trailer), "31" (Car/Sedan), "32" (Truck), "41" (Bus) and "42" (Big Truck) as shown at Figure 11 and Figure 12. For northern directions at national freeway 01 and national freeway 03 both distributions become busy start at 06.00 am and reached its peak at 10.00 am. Furthermore, both distributions also dominated by car/sedan vehicle.

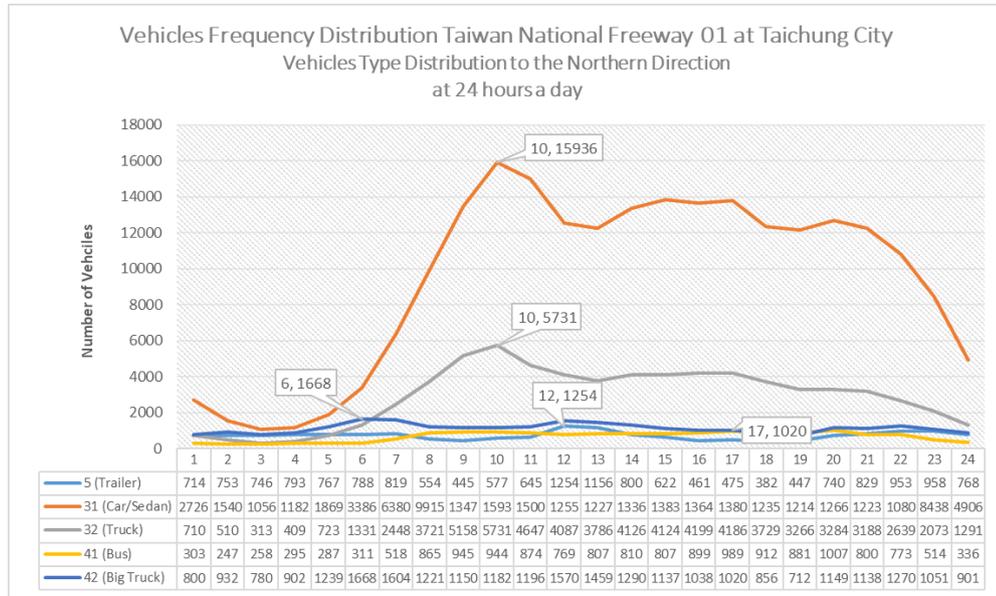

Figure 11. Traffic Distribution National Freeway 01 at Taichung city northern direction based on vehicle types

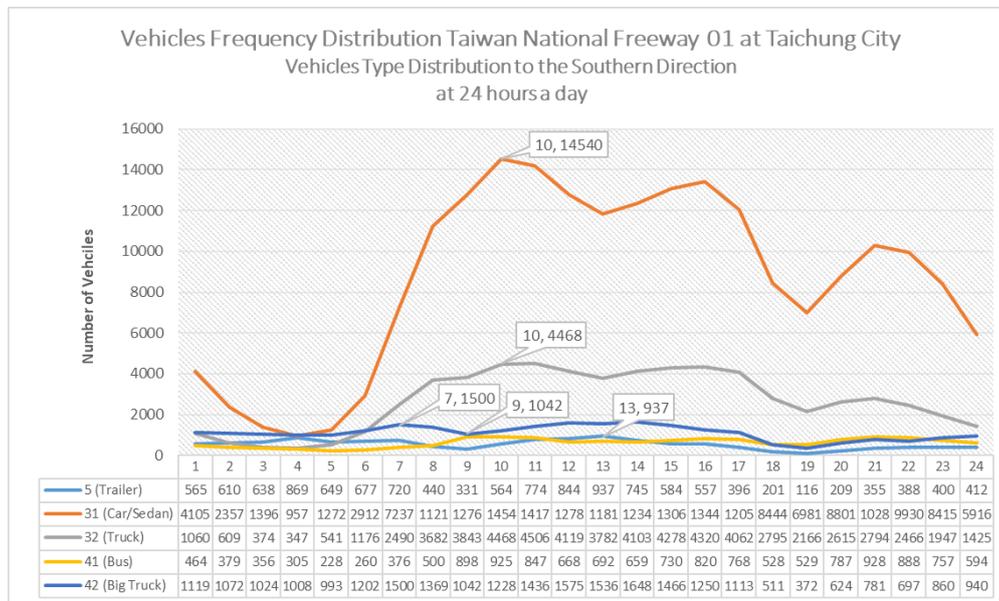

Figure 12. Traffic Distribution National Freeway 03 at Taichung city northern direction based

On the other hand, southern directions at national freeway 01 and national freeway 03 shown at Figure 13 and Figure 14 respectively. Both distributions become busy start at 14.00 am and reached its peak at 17.00 in the evening. Furthermore, both distributions also dominated by car/sedan vehicle.

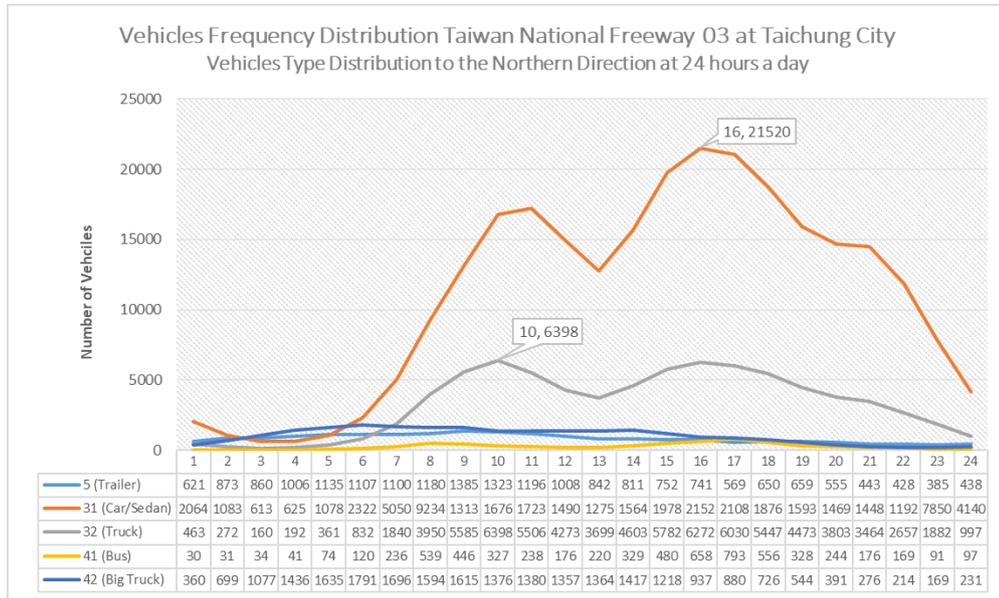

Figure 13. Traffic Distribution National Freeway 01 at Taichung city southern direction based on vehicle types

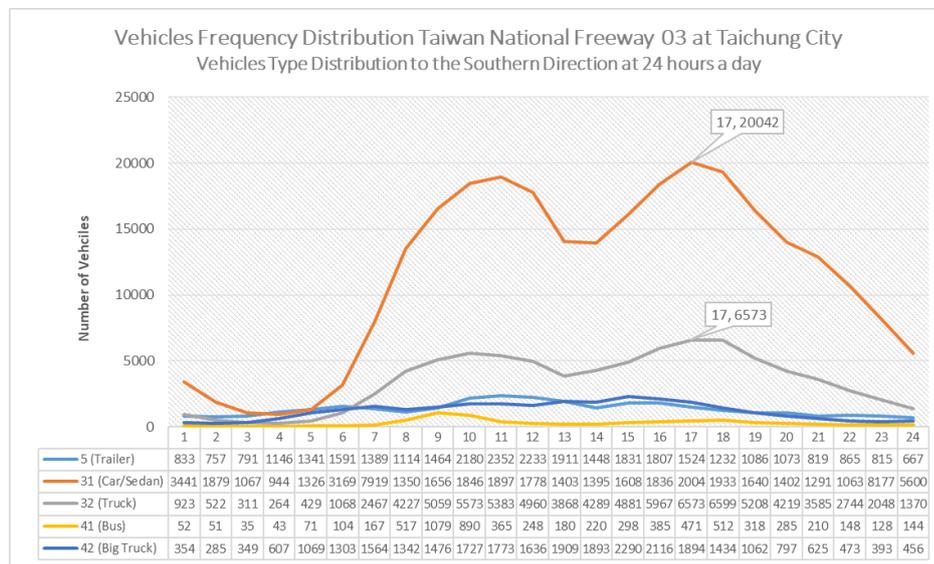

Figure 14. Traffic Distribution National Freeway 03 at Taichung city southern direction based on vehicle types

## 4.4. Vehicle Average Time Travel Traffic Distribution per week

Information about average travel time to northern or southern directions during weekend or weekday is interesting. People will know how long they will travel through that freeway. Figure 15 and Figure 16 shown average vehicles distribution to northern direction at national freeway 01 and national freeway 03 respectively. The average travel time ranging from 12 to 18 minutes. Average travel time reached its peaked during 13.00 to 16.00 in the afternoon.

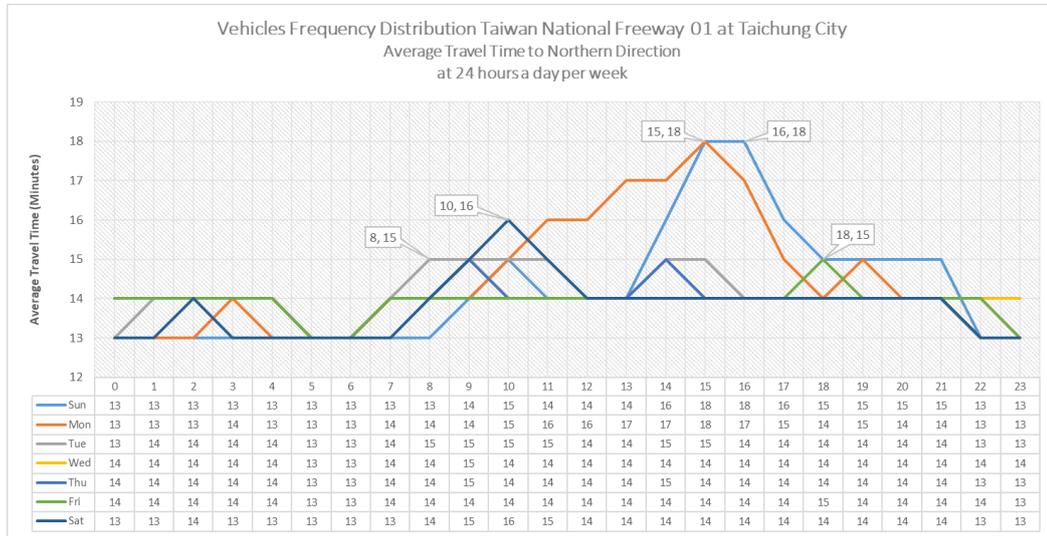

Figure 15. Traffic Distribution National Freeway 01 at Taichung city northern direction based on average time distribution

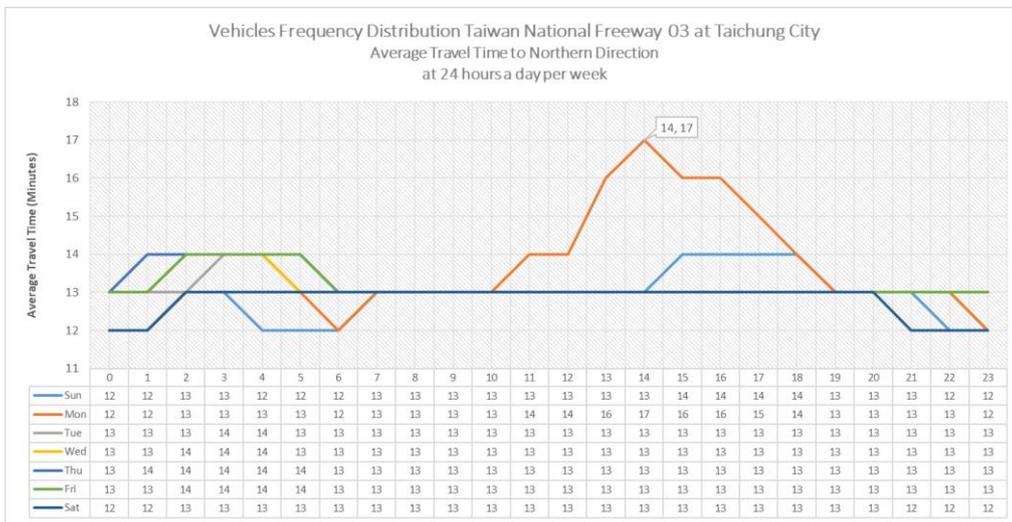

Figure 16. Traffic Distribution National Freeway 03 at Taichung city northern direction based on average time distribution

Furthermore, at Figure 17 and Figure 18 shows average travel time to southern directions. The average travel time ranging from 12 to 16 minutes. It is important to take a note that national freeway 01 the average travel time will peak from 16.00 to 18.00 for weekday and weekend. On the other hand, national freeway 03 the average travel time will peak from 12.00 to 14.00 in the afternoon.

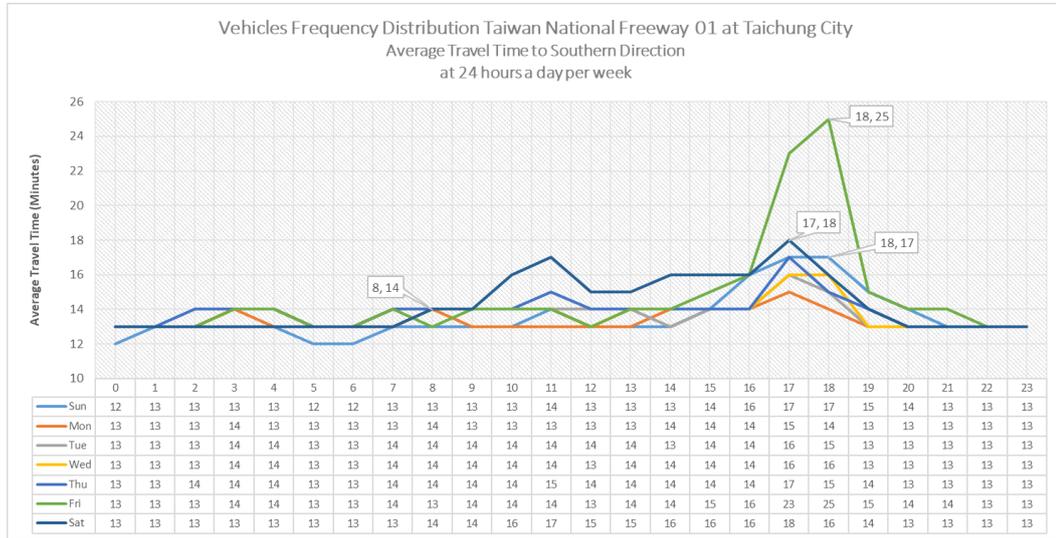

Figure 17. Traffic Distribution National Freeway 01 at Taichung city southern direction based on average time distribution

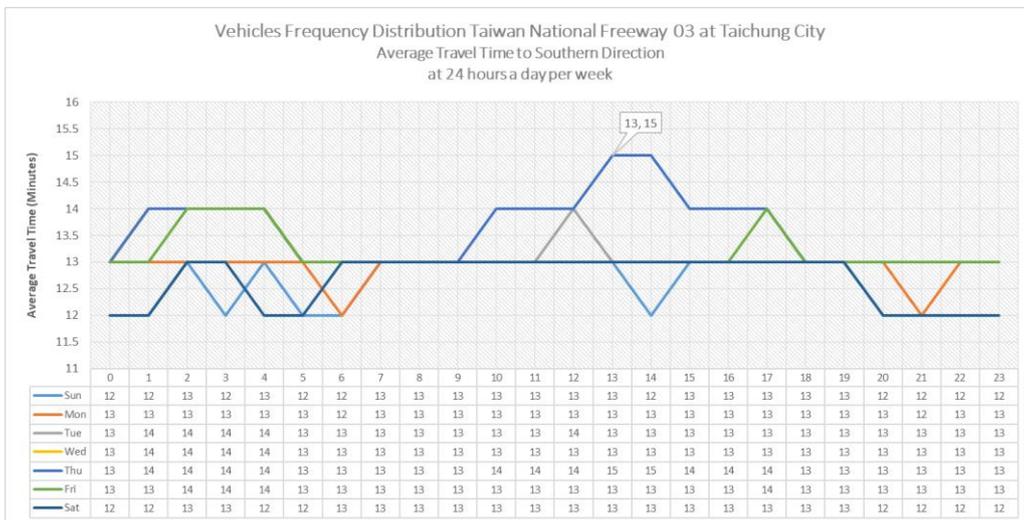

Figure 18. Traffic Distribution National Freeway 03 at Taichung city southern direction based on average time distribution

# 5. Conclusion and Recommendation

In this paper, we describe a method to extract travel pattern interval using Gantries freeway data from Traffic Data Collection System (TDCS) Taiwan open data. Experimental from one-month data at September 2018. Furthermore, Hadoop MapReduce framework can be utilized as to retrieve information at very large data effectively. In the future research, statistic information about industrial data, school data and tourism are can be included. These data could enhance significant information retrieved from the system.